\documentclass[conference]{IEEEtran}

\usepackage{multirow}
\usepackage{subfig}
\usepackage{graphicx}
\usepackage{booktabs} 
\let\labelindent\relax
\usepackage{enumitem}
\usepackage{changes}

\newcommand{\rowgroup}[1]{\hspace{-0.5em}\underline{#1}}
\ifCLASSINFOpdf
\else
\fi
\begin{document}
%
\title{Quantifying Users' Beliefs about Software Updates}

\author{\IEEEauthorblockN{Arunesh Mathur\IEEEauthorrefmark{1},
Nathan Malkin\IEEEauthorrefmark{2}, Marian Harbach\IEEEauthorrefmark{3},
Eyal Peer\IEEEauthorrefmark{4} and Serge Egelman\IEEEauthorrefmark{3}\IEEEauthorrefmark{2}}
\begin{tabular}{cc}
\IEEEauthorrefmark{1}\emph{Princeton University} & \IEEEauthorrefmark{2}\emph{University of California, Berkeley} \\
\IEEEauthorrefmark{3}\emph{International Computer Science Institute} & \IEEEauthorrefmark{4}\emph{Consumers Behavioral Insights Lab, Bar-Ilan University, Israel}
\end{tabular}
\IEEEauthorblockN{
\begin{tabular}{ccc}
amathur@cs.princeton.edu & nmalkin@cs.berkeley.edu & mharbach@icsi.berkeley.edu\\
\multicolumn{3}{c}{
eyal.peer@biu.ac.il \hspace{0.2cm} egelman@cs.berkeley.edu
}\\
\end{tabular} \\
}}


%


\maketitle

\begin{abstract}
Software updates are critical to the performance, compatibility, and security
of software systems. However, users do not always install updates, leaving
their machines vulnerable to attackers' exploits. While recent studies have
highlighted numerous reasons why users ignore updates, little is known about
how prevalent each of these beliefs is. Gaining a better understanding of the
prevalence of each belief may help software designers better target their
efforts in understanding what specific user concerns to address when developing
and deploying software updates. In our study, we performed a survey to
quantify the prevalence of users' reasons for not updating uncovered by
previous studies. We used this data to derive three factors underlying these
beliefs: update costs, update necessity, and update risks. Based on our
results, we provide recommendations for how software developers can better
improve users' software updating experiences, thereby increasing compliance
and, with it, security.

\end{abstract}


%

\section{Introduction}


Software updates are essential to keeping systems and programs
up-to-date. These updates fix bugs and bring about improvements in performance
and usability; but arguably their most important function is enhancing system
security by fixing vulnerabilities. In 2015 alone, Microsoft reported 3,300
vulnerability disclosures of varying threat levels and estimated that close to
a quarter of Windows Personal Computers (PCs) were not always protected and updated to the latest patch level~\cite{MS_Security_Intelligence_Report_2015}.
Similarly, Cisco suggested that most security exploits will continue to be
propagated by outdated software that contains known
vulnerabilities~\cite{Cisco_Intelligence_Report_2015}. Therefore, these
companies and security agencies---such as the United States Computer Emergency
Readiness Team~\cite{USCERT_TIP}---recommend that users install
software updates as soon as they become available, in order to protect systems
from being exploited by attackers. In fact, applying updates in a timely manner
is one of the few pieces of computer security advice on which
experts agree~\cite{Ion_2015_Noone_Can_Hack_My_Mind}.

However, recent studies have shown that users avoid or delay installing
software updates~\cite{Vaniea_CHI_2016, Vaniea_CHI_2014, Mathur_SOUPS_2016,
Tian_2015, Forget_SOUPS_2016, mathur2017impact} and uncovered some of the
reasons that users offer for this behavior. While these studies make a timely
contribution to aid software developers, they are primarily qualitative. As a
result, we know very little about the prevalence of these beliefs. Given that
beliefs shape behavior~\cite{ajzen1991theory}, understanding exactly how
widespread these beliefs are will help software developers and security
professionals more efficiently understand how to address users' concerns and
improve the overall software updating experience, and ultimately, affect security.


In our study, we conducted a survey to better understand how updating beliefs
vary among the general population, and investigate whether they can be grouped into actionable factors for the community.
We contribute to the literature on software updating behaviors by
quantifying the prevalence of various beliefs about software updates, as uncovered by the previous qualitative literature. Our
analysis uncovers three distinct factors that highlight why users claim to
avoid updates:

\begin{itemize}
\item\emph{Update Costs}: These costs include the time it takes to install the update, whether a restart is required, and its required space on disk.
\item\emph{Update Necessity}: This includes users' satisfaction with the current system or program, whether the update's purpose is clear, and its perceived importance.
\item\emph{Update Risks}: These risks include data loss due to the update, and whether the update may be malicious.
\end{itemize}

We interpret and discuss our results, highlighting broad implications these have for developers who design software updates for users. Understanding these unifying factors may help guide developers towards improving the updating experience through better software design and messaging.

\section{Related Work}




Failure to patch known security vulnerabilities is one of the leading causes for
security breaches; most exploits target systems that have not been patched,
rather than undisclosed zero-days~\cite{Cisco_Intelligence_Report_2015}.
This impacts both end-users and system administrators: for instance, the recent
Equifax data breach, the largest of its kind, was as a result of a system administrator's
failure to apply a software update~\cite{Newman2017}. This update
fixed a known vulnerability and had been available for several months.
Thus, ensuring that people update their systems in a timely fashion would help in improving computer security.

Only recently have security researchers begun exploring how users feel about and respond to software
updates. Ion \emph{et al.}\ \cite{Ion_2015_Noone_Can_Hack_My_Mind} compared the
advice expert and non-expert computer users gave in order to stay protected
online. They found that non-experts installed updates less frequently and
lacked awareness about the importance of updates. In a somewhat representative
sample of the US, Wash and Rader~\cite{Wash_Rader_2015_Too_Much_Knowledge}
found that only 24.2\% of users reported taking ``advanced'' security
actions---including installing software updates and patches---frequently. Traces of update habits of Android users have shown that only about 50\% of all users update to a new application version within the first week of the update's release~\cite{pinornot}. Software developers have also been shown to not update third-party libraries after first use~\cite{derr2017keep}.

A related set of studies have explored in greater depth why users avoid or
delay installing software updates. Rader and
Wash~\cite{Wash_2014_Out_Of_The_Loop_Automated_Updates_Consequences} found that
automatic updates that try to involve users in the decision-making process lead
to poor mental models of updating and result in less secure systems. Vaniea
\emph{et al.}\ \cite{Vaniea_CHI_2014} interviewed 37 Windows users and found
that they cited three reasons for avoiding updates: updates introduced
undesirable features, the value and purpose of an update was hard to assess,
and the need for updating was unclear because the software/program functioned
correctly. Mathur and Chetty~\cite{mathur2017impact} found that users who have
negative experiences with software updating disabled auto-updates on Android
devices.

In a survey of 155 users, Fagan \emph{et al.}\
\cite{Fagan_2015_Users_Software_Update_Messages} explored software update
notifications and found design features that led to annoying and confusing
messages. Mathur \emph{et al.}\ \cite{Mathur_SOUPS_2016} interviewed 30 users
and found they felt annoyed with notifications that interrupted their tasks,
including having to restart their machines. They described software updating as
an ``information problem,'' highlighting the different pieces and sources of
information users sought before updating, including the source of the update
and the duration of the install process. Forget \emph{et al.}\
\cite{Forget_SOUPS_2016} compared the level of end-user engagement in the
management of their computers' security with security outcomes and discovered
that greater engagement did not always correspond to greater security. The
authors uncovered a variety of reasons why users avoid or delay updates,
confirming the results of previous studies~\cite{Vaniea_CHI_2014,
Vaniea_CHI_2016, Wash_2014_Out_Of_The_Loop_Automated_Updates_Consequences}.
They suggested that security mitigations, such as software updating, need to be
designed according to how much users engage with computer security. Vaniea and
Rashidi~\cite{Vaniea_CHI_2016} surveyed 307 users about their experiences with
updating software and highlighted various user experiences at each step of the
process.

Our study builds on this previous work, which uncovered users' underlying beliefs about updating their software, by measuring the prevalence of these various beliefs. Our study uses this to offer practical recommendations for software developers to consider when developing and designing updates in order to deliver a better user experience.

\section{Method}


\subsection{Survey Construction and Deployment}

We first surveyed the literature on why users avoid software updates, noting each belief we came across. Not only does the existing literature provide a breadth of users' beliefs about avoiding updates, but also validates it with actual behavior~\cite{Forget_SOUPS_2016,Vaniea_CHI_2014}---that is, users leaving their systems unpatched---in many studies. We discovered a total of 15 different beliefs relating to software updates:

\begin{itemize}[wide,leftmargin=*]
\item [B1-Fine:] Updates seem unnecessary because everything works fine~\cite{Vaniea_CHI_2014, Tian_2015, Vaniea_CHI_2016}
\item [B2-Time:] Updates take a long time to install~\cite{Mathur_SOUPS_2016, Vaniea_CHI_2016}
\item [B3-Restart:] Updates require unnecessary restarts of applications or PCs~\cite{Wash_2014_Out_Of_The_Loop_Automated_Updates_Consequences, Mathur_SOUPS_2016, Forget_SOUPS_2016, Vaniea_CHI_2016}
\item [B4-Unused:] Updates are requested by programs used infrequently~\cite{Vaniea_CHI_2014, Vaniea_CHI_2016, Mathur_SOUPS_2016, Forget_SOUPS_2016}
\item [B5-Unimportant:] Updates seem unimportant~\cite{Mathur_SOUPS_2016, Forget_SOUPS_2016, Vaniea_CHI_2014}
\item [B6-Purpose:] The purpose of updates is unclear and hard to understand~\cite{Mathur_SOUPS_2016}
\item [B7-Bundled:] Updates introduce unwanted bundled programs~\cite{Vaniea_CHI_2016}
\item [B8-Features:] Updates add unwanted or remove wanted features from programs~\cite{Vaniea_CHI_2016}
\item [B9-Bugs:] Updates introduce new bugs into programs~\cite{Tian_2015, Mathur_SOUPS_2016}
\item [B10-Space:] Updates occupy a lot of disk space~\cite{Mathur_SOUPS_2016, Vaniea_CHI_2016}
\item [B11-Compatibility:] Updates cause compatibility issues between programs~\cite{Mathur_SOUPS_2016, Vaniea_CHI_2016}
\item [B12-UI:] Updates disrupt program user interfaces~\cite{Vaniea_CHI_2014, Vaniea_CHI_2016, Mathur_SOUPS_2016, Forget_SOUPS_2016}
\item [B13-DataCost:] Updates consume a lot of data on Internet plans~\cite{Mathur_SOUPS_2016}
\item [B14-Malicious:] Updates contain malicious software~\cite{Mathur_SOUPS_2016, Vaniea_CHI_2016}
\item [B15-DataLoss:] Updates lead to a loss of data~\cite{Vaniea_CHI_2016}

\end{itemize}

We compiled these beliefs into a survey where we asked participants how often,
in their experience, each one of these statements about software updates were
true. (Respondents rated their beliefs on a five-point scale, from ``rarely'' to
``always.'')
We chose not to ask about
any specific updating experience
(such as specific devices or operating systems),
to avoid bias from any recent or otherwise memorable experience.
We wanted to collect opinions that people have come to hold about software updates in general,
as these would color users' interactions with any new software update.
In addition to the ratings,
we collected participants' demographic information: age, gender, and (if available)
occupation.

We recruited participants from Amazon Mechanical Turk (AMT),
an online work marketplace that
has been shown to produce diverse and generally representative samples of the population~\cite{buhrmester_amazons_2011}.
We limited the
survey to only those AMT users who were based in the United States (US) and had
a completion rate of 95\% or greater. Sampling high-reputation AMT users ensured we did not have to resort to attention check questions~\cite{peer2014reputation}. The survey took between 5-10 minutes, and
participants were compensated \$2 for their time. The beliefs were randomized for each AMT user to avoid any potential ordering bias.

We also launched the survey
on Google Consumer Surveys (GCS) to compare the prevalence of these beliefs
across two samples. Because the number of questions in our survey (15) exceeded
the maximum number of questions GCS allows in a survey (10), we placed each
question in a survey by itself. GCS participants were compensated with Play
Store credit. 
This study was approved by our Institutional
Review Board.

\begin{table}[ht]
  \small
  \centering
  \begin{tabular}{l r r}
\toprule
        \small{\textbf{Demographic}}  &
    \small{\textbf{AMT Participants}} & \small{\textbf{GCS Participants}}\\ 
    
    \midrule
    \rowgroup{\small{\textbf{Age}}} & \\
    
        18--24 & 2.49\% & 14.58\%\\
        25--34 & 40.80\% & 20.56\%\\
        35--44 & 30.85\% & 15.54\%\\
        45--54 & 11.94\% & 12.47\%\\
        55--64 & 8.46\% & 11.70\%\\
        \textgreater= 65 & 5.47\% & 8.57\%\\
        
    \midrule
    
    \rowgroup{\small{\textbf{Gender}}} & \\
    Male & 58.71\% & 42.97\%\\
    Female & 41.29\% & 42.23\%\\
    Other & 0.00\% & 14.8\%\\
    \midrule
    
    \rowgroup{\small{\textbf{Education}}} & \\
                         12th grade or less & 1.49\% & NA\\
                         High school & 12.44\% & NA\\
                         Some college & 17.91 & NA\\
                         Bachelor's & 45.27\% & NA\\
                         Associate's & 12.94\% & NA\\
                         Post-graduate & 9.45\% & NA\\
                         Prefer not to answer & 0.50\% & NA\\
    \bottomrule

  \end{tabular}
  \caption{Demographic Information of the Amazon Mechanical Turk (AMT) and Google Consumer Survey Participants (GCS). GCS does not report data about participants' education levels.}
  \label{tab:demographics}
\end{table}

\subsection{Data Analysis}

The beliefs we compiled cover a wide range of the updating process's aspects,
and we wanted to uncover any underlying themes that potentially unified multiple beliefs.
Rather than grouping the beliefs thematically,
based on our own subjective judgments,
we decided to perform principal component analysis (PCA)~\cite{hotelling1933analysis} with Varimax rotation 
on the responses of participants from our AMT sample. This technique has previously been used extensively in psychology and human-computer interaction literature~\cite{pca1,pca2,pca3,pca4}, as it yields more descriptive factors which are robust to correlation. However, we also confirmed our factor loadings using an Oblimin rotation.
The resulting dimensionality reduction
helps unveil which beliefs are frequently held together,
and allows us
to explore the factors underlying why users avoid updating their
software.

Because we were interested in factors resulting from the prevalence of a belief (and not its strengths), we
transformed the original belief scores---measured on a 5-point scale---to a binary variable for our analysis. We encoded
``never (1),'' ``rarely (2),'' and ``sometimes (3)'' as 0---indicating that
users did not hold that belief, and encoded ``often (4)'' and ``always (5)'' as
1---indicating that users held that belief.

\section{Findings}


We received 200 complete responses from AMT and, on average, close to 200 responses for each belief on GCS.
Our participants were roughly balanced in gender and age;
detailed demographics can be seen in Table~\ref{tab:demographics}.

\begin{figure*}[th]
    \centering
    \subfloat[Amazon Mechanical Turk]{
        \includegraphics[width=.50\textwidth]{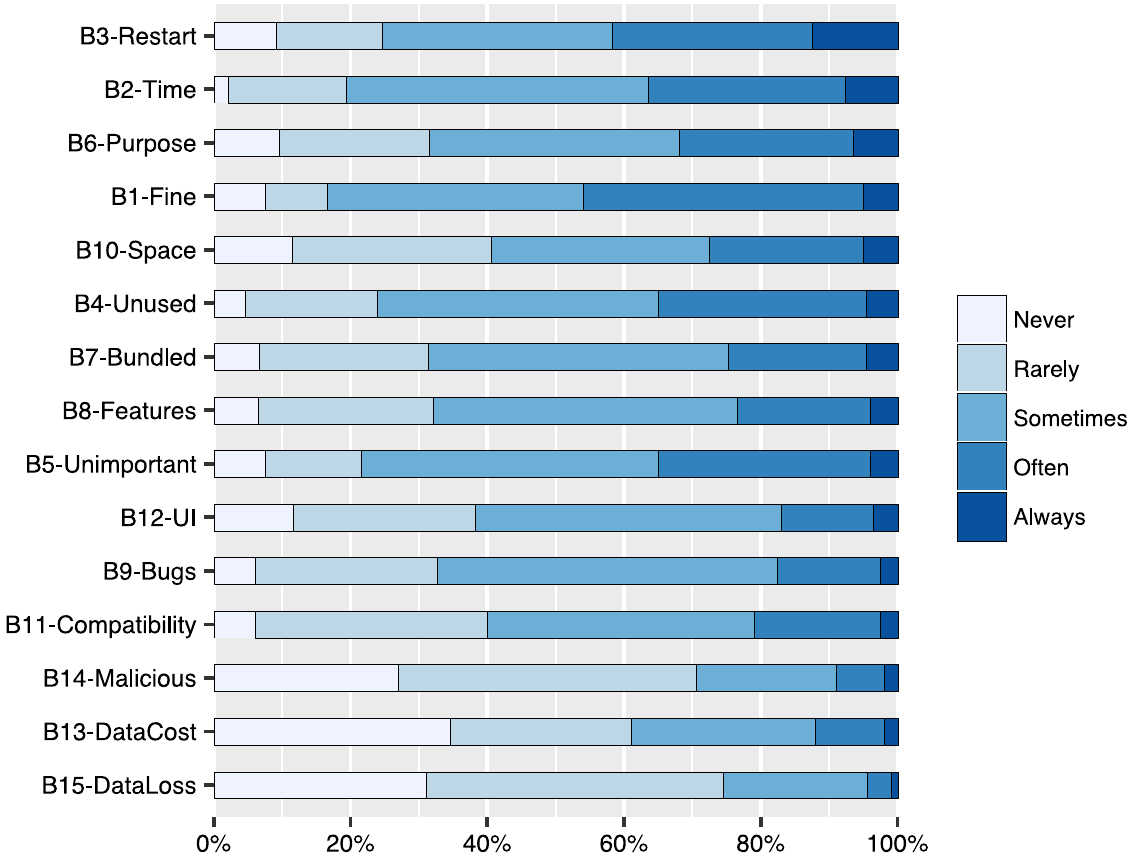}
        \label{fig:AMT}
    }
    \subfloat[Google Consumer Surveys]{
        \includegraphics[width=.50\textwidth]{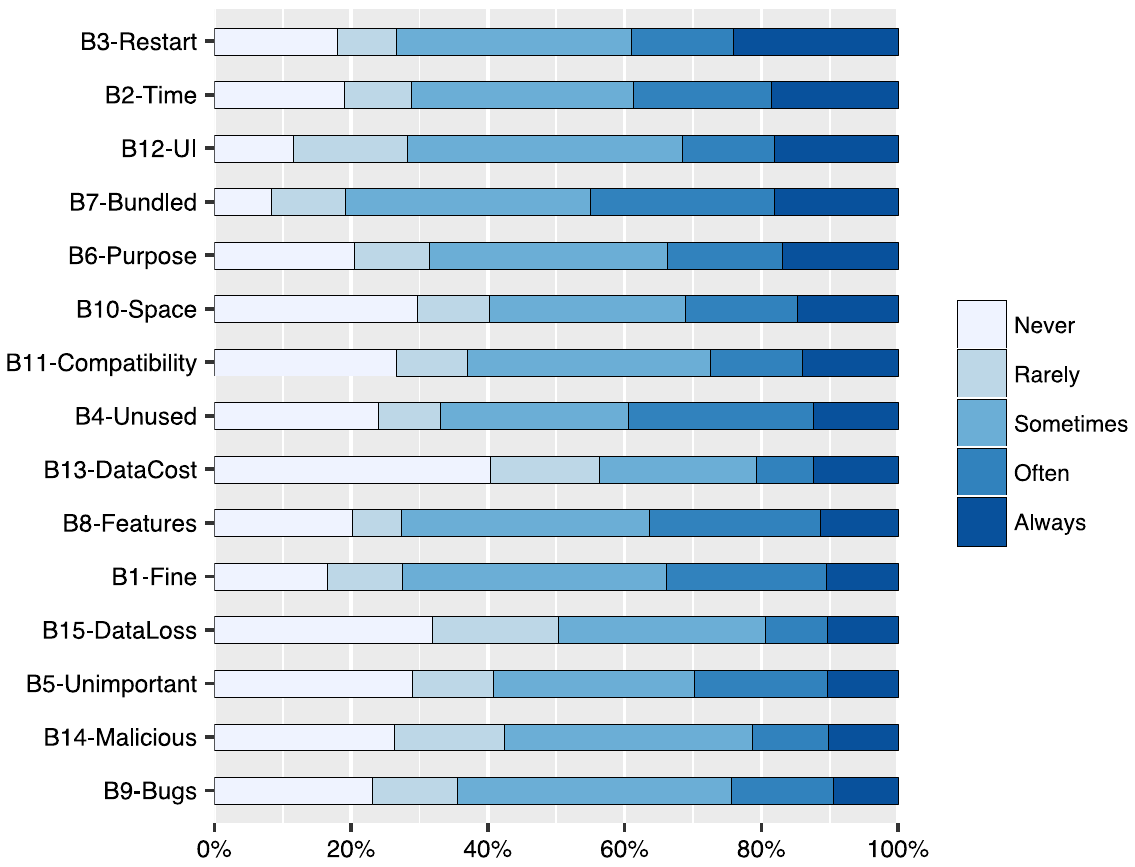}
         \label{fig:GCS}
    }
    
        \caption{The distribution of software updating beliefs from both the Amazon Mechanical Turk and Google Consumer Surveys samples.} 
    \label{fig:sample}
\end{figure*}

\subsection{Comparing the Two Samples}

\begin{figure}[th]
    \centering
        \includegraphics[width=0.45\textwidth]{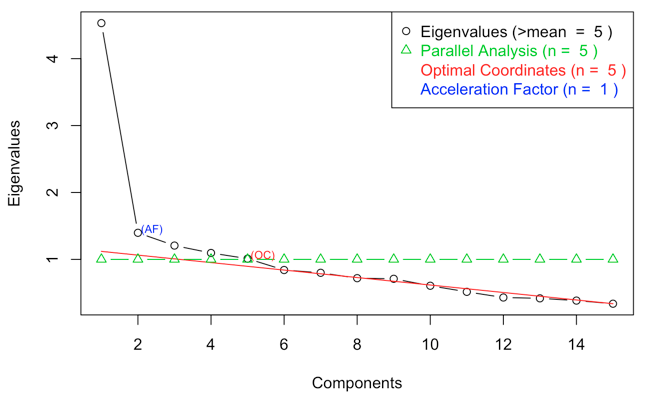}    
        \caption{Scree plot indicating the number of factors we discovered during the Exploratory Factor Analysis.}
    \label{fig:scree}
\end{figure}

We first compared the AMT and GCS responses, conducting a Mann-Whitney U test
to examine whether participants' beliefs differed across samples; because we conducted 15 such tests, we applied the Bonferroni correction, and therefore only report significance at $p < 0.003$. The distributions of the beliefs across the two samples are presented in Figure~\ref{fig:sample}.

We found that both samples were mostly in agreement with respect to their
beliefs. The differences between AMT and GCS responses were statistically
significant for four of the software updating beliefs [B7 ($r = 0.22$), B12 ($r
= 0.17$), B14 ($r = 0.20$), B15 ($r = 0.17$)], with small ($r = 0.1$) to medium
($r = 0.3$) effect sizes~\cite{coolican1990research}. Participants in the GCS
sample had a stronger belief that:

\begin{itemize}
\item updates introduce unwanted bundled programs into software (B7),
\item updates disrupt the user interface of programs (B12),
\item updates contain malicious software (B14), and
\item updates lead to a loss of data (B15).
\end{itemize}

For all other beliefs, we found no significant differences in prevalence between the two samples.

As described above,
the limitations of the GCS platform required us to ask about each belief in a separate survey.
These surveys are anonymous,
preventing us from examining beliefs across individuals.
The GCS sample therefore serves primarily as evidence of the AMT sample's external validity,
and we focus the remainder of our analysis on our AMT participants. (Prior work has already shown
that results from AMT are consistent with those of laboratory experiments~\cite{Paolacci2010,horton2011online},
and that AMT samples ``produce reliable results with standard decision-making biases''~\cite{Goodman2013}.)


\subsection{Exploratory Factor Analysis}

In our dataset, we found five eigenvalues greater than 1.0 (the Kaiser
criterion~\cite{Fabrigar1999}), suggesting the presence of up to five factors. We also verified the presence of these five factors using a scree plot (see Figure~\ref{fig:scree}). Together, these five factors explained 62\% of the variance in the data.
Subsequently, we extracted these five factors using a PCA, applied a Varimax rotation and considered a belief loaded on a
factor if its loading exceeded 0.5. We ignored an item if it loaded on a
factor, but its loading on that factor was not twice as high as its loading on
the other factors~\cite{saucier1994mini}. These factor loadings are presented
in Table \ref{tab:table2}.


After this step, we ignored beliefs B4, B7, B8, B9, B11, and B12, as they
failed to load predominantly on a single factor.
We note that this does not reflect the importance of these beliefs,
but only that they are not strongly correlated
or that they are held at equal rates by the population.
This is a necessary step for PCA,
providing a more reliable basis to draw conclusions from,
as leaving in individual questions would cause too much bias.

Left with the remaining 9
beliefs, we repeated the PCA, Varimax rotation, and computed the new loadings.
We extracted three factors explaining 56\% of the variance in the data. The
output of this step is shown in Table \ref{tab:table3}.
We emphasize again
that we could not repeat this analysis on the GCS sample since those participants
were not the same across the GCS survey questions (i.e., each GCS participant
only answered a single question).

\begin{table}[ht]
\small
  \centering
  \begin{tabular}{l c c c c c}
  \toprule
  \small{\textbf{Belief}} & \small{\textbf{Fac.\ 1}} & \small{\textbf{Fac.\ 2}} & \small{\textbf{Fac.\ 3}} & \small{\textbf{Fac.\ 4}} & \small{\textbf{Fac.\ 5}} \\
  \midrule
  B3-Restart & 0.651 & & & & \\ 
  B8-Features & 0.620 & & 0.363 & & \\ 
  B7-Bundled & 0.582 & & 0.397 & & \\ 
  B2-Time & 0.581 & & & & \\ 
  B4-Unused & 0.509 & & & 0.345 & \\
  B9-Bugs & 0.455 & & 0.452 & \\
  B1-Fine & & 0.840 & & & \\
  B5-Unimport. & & 0.794 & & & \\
  B6-Purpose & & 0.629 & & & \\
  B14-Malicious & & & 0.785 & & \\
  B15-DataLoss & & & 0.761 & & \\
  B12-UI & 0.407 & 0.357 & 0.461 & & \\
  B13-DataCost & & & & 0.866 & \\
  B11-Compatib. & & & & 0.551 & 0.524 \\
  B10-Space & & & & & 0.790\\
   \bottomrule
  \end{tabular}
  \caption{Factor loadings from the first PCA with Varimax rotation. Only loadings \textgreater 0.25 are shown.}~\label{tab:table2}
\end{table}

\begin{table} [ht]
\small
  \centering
  \begin{tabular}{l c c c}
  \toprule
  \small{\textbf{Belief}} & \small{\textbf{Necessity}} & \small{\textbf{Costs}} & \small{\textbf{Risks}} \\
  \midrule 
  B1-Fine & 0.779  & & \\
  B5-Unimportant & 0.743 & 0.318 & \\
  B13-DataCost & 0.546 & & \\
  B2-Time & & 0.745 & \\
  B3-Restart & & 0.709 & \\
  B10-Space & & 0.536 & \\
  B14-Malicious & & & 0.819 \\
  B15-DataLoss & & & 0.810 \\
   \bottomrule
  \end{tabular}
  \caption{Factor loadings from the second PCA with Varimax rotation, after removing beliefs that did not load at least twice as highly on a single factor during the previous PCA. Only loadings \textgreater 0.25 are shown.}~\label{tab:table3}
\end{table}

The process left us with three distinct factors,
each representing 2--4 beliefs.
We gave each of the factors names
based on the ideas that unify the beliefs that loaded onto them:

\begin{itemize}
\item {\it Update Necessity}: These beliefs concern failing to understand why updates are required and what purpose they serve:
	\begin{itemize}
	\item B1 (``Updates seem unnecessary as everything works fine'')
	\item B5 (``Updates seem unimportant'')
	\item B13 (``Updates consume a lot of data on Internet plans'')
	\end{itemize}
\item {\it Update Costs}: These beliefs represent the costs users incur while updating (whether they be time, resources, etc.):
	\begin{itemize}
	\item B2 (``Updates take a long time to install'')
	\item B3 (``Updates require unnecessary restarts of applications or PCs'')
	\item B10 (``Updates occupy a lot of disk space'')
	\end{itemize}
\item {\it Update Risks}: These beliefs represent risks faced during or after updating:
	\begin{itemize}
	\item B14 (``Updates contain malicious software'')
	\item B15 (``Updates lead to a loss of data'')
	\end{itemize}
\end{itemize}

\subsection{Prevalence of Beliefs}

We calculated the percentage of AMT participants whose beliefs were described by each of the factors: for the beliefs comprising each factor, we took the average Likert score, and then applied the same threshold we used for PCA. In this manner, we observed that 40.5\% of participants held beliefs about \emph{Update Costs}, 29.2\% held beliefs about \emph{Update Necessity}, and 7.5\% held beliefs about \emph{Update Risks}.

To examine discriminant validity between the three factors, we performed a Spearman correlation. We observed that while the three factors were inter-correlated, these correlations are not very strong: $r=0.478$ when comparing \emph{Updating Costs} with \emph{Updating Risks}, $r=0.487$ when comparing \emph{Updating Necessity} with \emph{Updating Risks}, and $r=0.597$ when comparing \emph{Updating Necessity} with \emph{Updating Costs}. Thus, participants may have beliefs relating to one factor, but not others. 

\subsection{Limitations}

A limitation of our study is that we did not follow-up our exploratory factor analysis with a confirmatory one.
Future research could replicate our analyses and validate the factors we found. Our results are also generalizable only to the AMT population. However, we supplemented our sample from AMT with another from GCS to provider a comparison of the beliefs. Further, while the AMT population is limited in how diverse it is, research \cite{bartneck2015comparing,simons2012common} has shown that it is similar to university students and other online participant pools.
Future studies may also wish to replicate our results with larger sample sizes.
We note, however, that we verified that our participant-to-item ratio (200/15 = 13:1)
was within the bounds considered sufficient for exploratory factor analysis
as shown by a meta study~\cite{costello2005best}.

\section{Discussion}


The ultimate goal of any research on software updates
is to make users more secure by increasing the number of out-of-date systems that get patched. Our research contributes to this goal by increasing our community's understanding
of the reasons why people do not update their systems.

\subsection{Software Updating Beliefs}
Our study is the first, to our knowledge, to quantify the frequency with which various beliefs about updating appear in the population. We found that most people expressed agreement with nearly all of the reasons for avoiding software updates uncovered by previous studies, to varying degrees (see Figure~\ref{fig:sample}). Except for narrow concerns (e.g., platform-specific), such as mobile data usage, and ones whose dangers have not been widely advertised (updates containing malicious software), the majority of respondents said each reason was at least sometimes true.

Respondents found certain aspects of the updating process especially annoying. About 40\% of the AMT and GCS sample stated that updates either ``often'' or ``always'' required unnecessary restarts. Similarly, the duration of the installation process was another frequent complaint, expressed by nearly 40\% of both the AMT and the GCS respondents, who said that this was either ``often'' or ``always'' a problem.

\subsection{Factors from Beliefs}
In addition to the prevalence of individual beliefs, our analysis uncovered several unifying factors. This finding suggests that, rather than holding an assortment of distinct and disconnected opinions, people subscribe to ``packages'' of beliefs about software updating. These represent three distinct axes along which users' concerns are expressed.

The first is \emph{Update Necessity}. Many respondents were unconvinced of the need for updates, agreeing that they ``seem unimportant'' because ``everything works fine.'' An orthogonal issue was the toll updates impose on users, which we refer to as \emph{Update Costs}. They take time and disk space,
and interrupt users' workflow when they demand a restart. Finally, some worried about \emph{Update Risks}, bad things that may happen if an update goes wrong: lost data or worse.

\subsection{Implications for Software Developers}

\subsubsection{Update Messaging}
Our results suggest that one relatively low-cost way of getting users to install updates is by solving the information problem,
i.e., accompanying updates with better messages. Currently, software updates typically present the same generic information about themselves to every user. These often lack concrete information and do not directly address users' inherent
reservations and beliefs about updating.

The dominant factors identified by our study suggest concrete topics update messaging can address.
For example, to address beliefs about \emph{Update Necessity}, developers can clarify the purpose specific updates serve,
explaining that they may be important even if everything is working fine.
This point is especially critical to make in the case of security updates, which
often fix problems that users are unlikely to even be aware of.

As another example, to address beliefs about \emph{Update Costs},
developers can clarify how users might be impacted
(e.g., the duration of the installation process) during the time of the update, 
and what if any, steps they need to take before installing the update for their specific operating system.
Conversely, if an update does not require a cost to end-users and runs in the background, the update message can clarify that. Future work could test these messages by means of controlled experiments.

\subsubsection{Designing and Deploying Updates}
Beyond simply messaging, developers should take these factors into account 
when designing software and updating mechanisms.
Thus, as the necessity of restarting is a common complaint,
and a major contributor to \emph{Update Costs},
systems that apply updates silently and in the background
would be welcomed by users.

Developers can take further steps to reduce their users' \emph{Update Risks}.
By using secure and verifiable updating mechanisms, they can reduce the possibility of malicious updates.
And by rigorous testing, the risk of data loss can be reduced.

Many modern systems have turned to automatic updates, which has has proven to be an effective and successful system~\cite{Nappa:2015:ACS:2867539.2867680}.
In cases where this is not an option,
it is to the benefit of both parties
for software developers to be able to convince their users to 
install updates.
For this to happen, a solid understanding of people's current beliefs, such as that established by this study,
is a fundamental prerequisite.

\section{Conclusion}

We used a large scale survey to measure the prevalence of different beliefs
users have about software updates that were discovered by previous qualitative literature~\cite{Vaniea_CHI_2014, Tian_2015, Vaniea_CHI_2016,Wash_2014_Out_Of_The_Loop_Automated_Updates_Consequences, Mathur_SOUPS_2016, Forget_SOUPS_2016}. We found that these beliefs can be grouped
into three factors that each represent a different facet of the beliefs:
\emph{Update Necessity}, \emph{Update Costs}, \emph{Update Risks}. These factors provide several practical recommendations for how software developers can improve existing update systems, including how and which of these previously discovered beliefs should be targeted and addressed.

\section*{Acknowledgements}
This research was supported by NSF grant CNS-1528070, BSF grant \#2014626 and The Center for Long-Term Cybersecurity at the University of California, Berkeley.



\bibliographystyle{IEEEtranS}
\bibliography{IEEEabrv,sample}
%

\end{document}